%% file: pandolfi.tex
\documentclass[10pt]{article}
\usepackage{graphicx}

\def\Title#1{\begin{center} {\Large #1 } \end{center}}
\def\Author#1{\begin{center}{ \sc #1} \end{center}}
\def\Address#1{\begin{center}{ \it #1} \end{center}}

\newcommand\pubblock{\rightline{\begin{tabular}{l} Proceedings of the Fifth Annual LHCP\\ \pubnumber\\
         \pubdate  \end{tabular}}}

\newenvironment{Abstract}{\begin{quotation} \begin{center} 
             \large ABSTRACT \end{center}\bigskip 
      \begin{center}\begin{large}}{\end{large}\end{center} \end{quotation}}

\newenvironment{Presented}{\begin{quotation} \begin{center} 
             PRESENTED AT\end{center}\bigskip 
      \begin{center}\begin{large}}{\end{large}\end{center} \end{quotation}}

\input econfmacros.tex

\usepackage{color}

\newcommand\pubnumber{ }

\newcommand\pubdate{\today}

\def\affiliation{
On behalf of the ATLAS and CMS Collaborations \\
}

\begin{document}

\large
\begin{titlepage}
\pubblock

\vfill
\Title{  Searches for New Heavy Resonances in Final States with Leptons and Photons in ATLAS and CMS  }
\vfill

\Author{ Francesco Pandolfi  }
\Address{\affiliation}
\vfill
\begin{Abstract}

Searches for resonances in final states with leptons and photons have always been a powerful tool for discovery in high energy physics. We present here the latest results from the ATLAS and CMS experiments, based on up to 36.1~fb$^{-1}$ of 13~TeV proton-proton collisions produced at the Large Hadron Collider. Detailed results on single lepton, dilepton, diphoton and Z$\gamma$ resonances are included.
\end{Abstract}
\vfill

\begin{Presented}
The Fifth Annual Conference\\
 on Large Hadron Collider Physics \\
Shanghai Jiao Tong University, Shanghai, China\\ 
May 15-20, 2017
\end{Presented}
\vfill
\end{titlepage}
\def\thefootnote{\fnsymbol{footnote}}
\setcounter{footnote}{0}
%

\normalsize 


\section{Introduction}

Resonances in leptons and photons have historically been a powerful tool for discovery in high energy physics. Notable examples come to mind, such as the discovery of the Z boson through its dielectron decay, by the UA1 and UA2 collaborations~\cite{ua1_w, ua1_z, ua2_w, ua2_z}, or the more recent discovery of the Higgs boson by the ATLAS~\cite{atlas} and CMS~\cite{cms} collaborations~\cite{higgsatlas, higgscms}, through its decay to photons or four leptons. The reason why these channels offer power tools for discovery is the high resolution with which electrons, muons and photons are reconstructed in typical high-energy particle detectors, and the relative low level of backgrounds that affect these searches.

Given the limited time allotted to this contribution, some choices necessarily had to be made: only a selection of results will be presented. The selection is largely personal, and it was my intent to favor recent results, particularly those which are based on the full 2016 LHC dataset of about 36 fb$^{-1}$ of 13~TeV proton-proton collisions. This contribution will therefore focus on the ATLAS results for searches in lepton + missing energy channels, the ATLAS results for searches of dielectron and dimuon resonances, the CMS results on searches for diphoton resonances, and CMS results for Z$\gamma$ searches.

\begin{figure}[t!]
\centering
\includegraphics[width=0.42\textwidth]{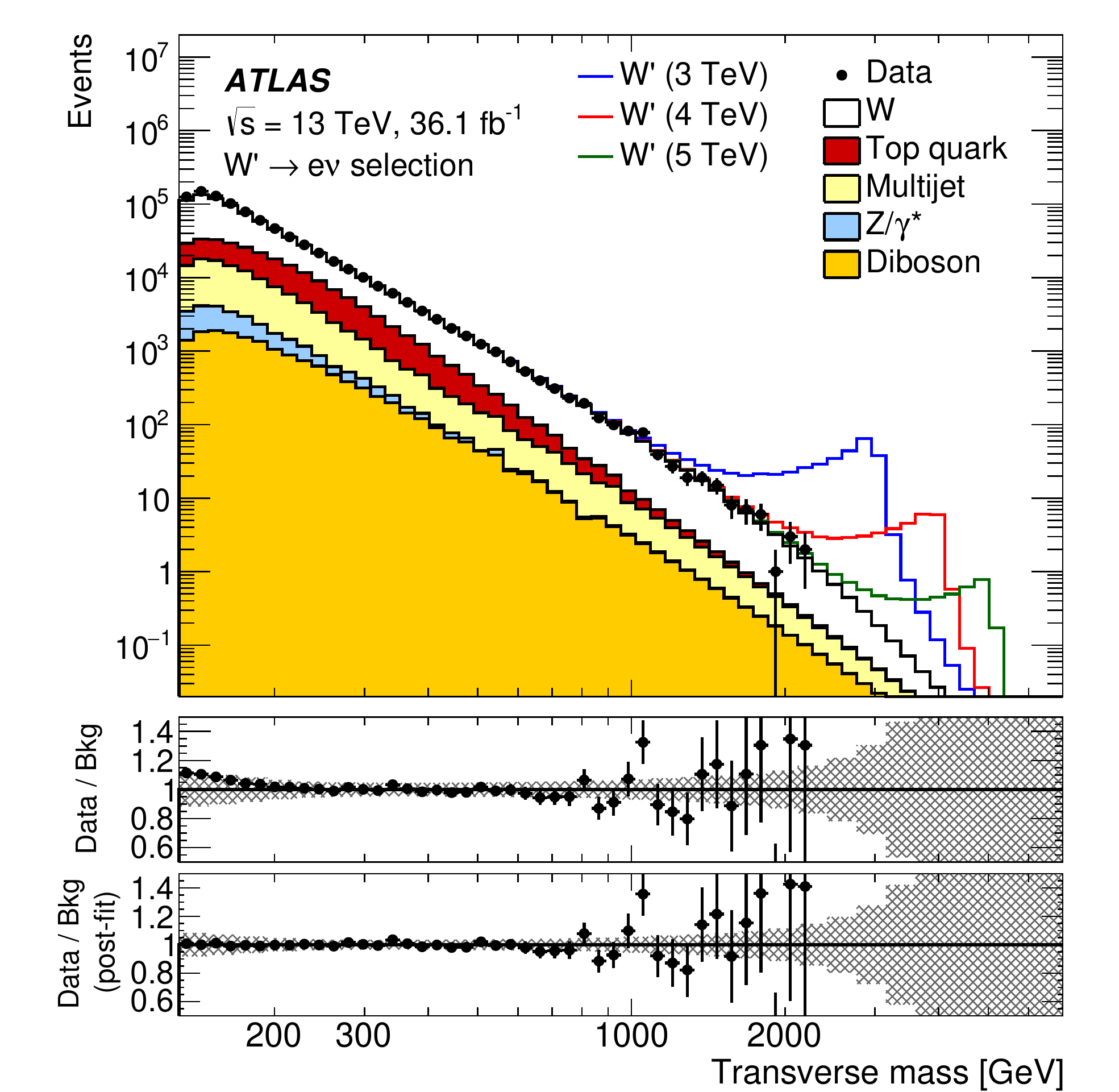}
\includegraphics[width=0.42\textwidth]{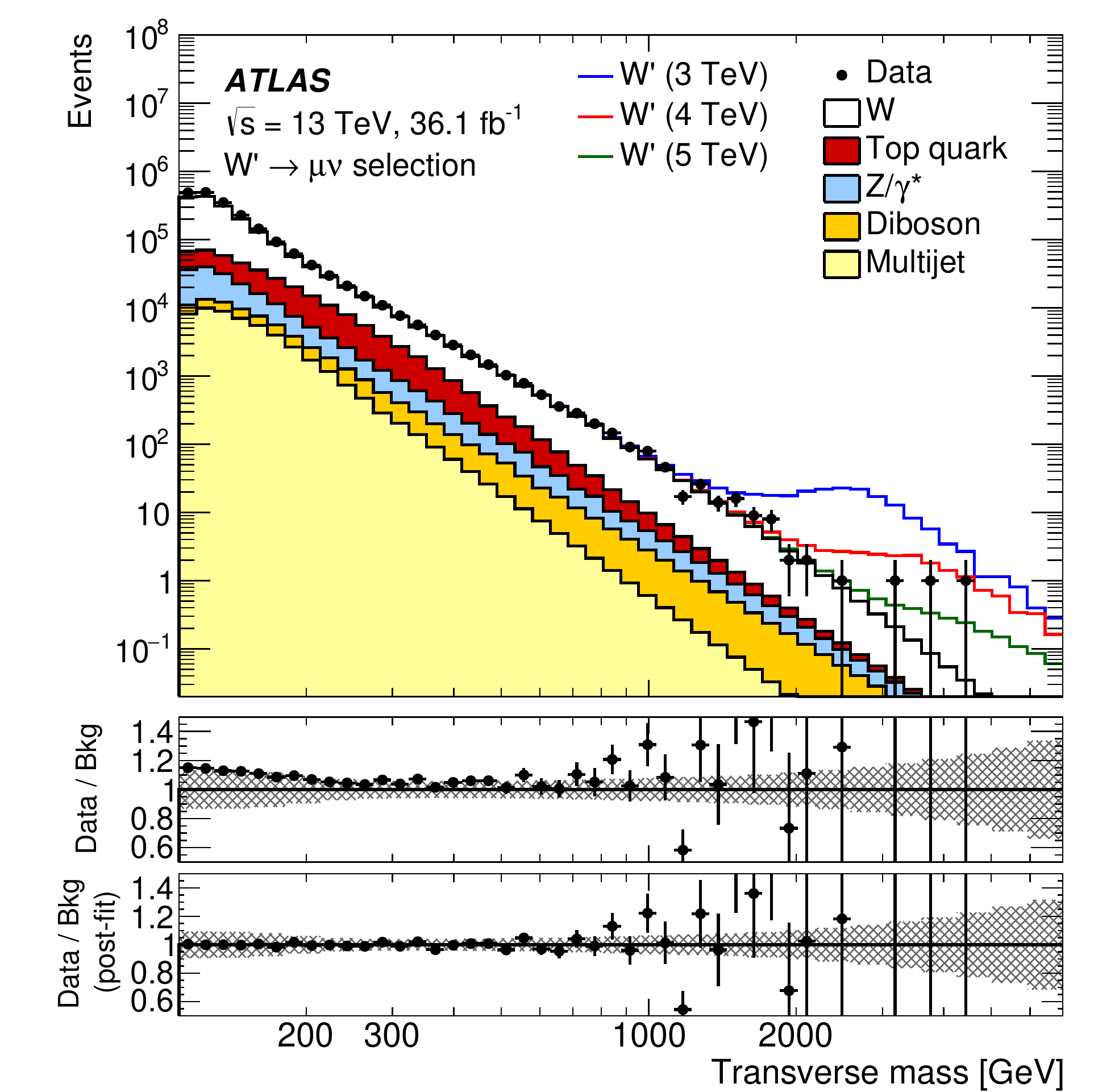}
\caption{Transverse mass spectra of the ATLAS single lepton searches in the electron~(left) and muon~(right) channels. The markers represent the observed spectra on 36.1~fb$^{-1}$ of data, while the stacked histograms represent the background estimates. Overlaid are shown also the expected shapes for benchmark Sequential Standard Model W$^{\prime}$ signals of 3, 4, and 5~TeV mass. The bottom pads show the ratio of the data to the sum of all backgrounds before (upper pad) and after~(lower pad) the background fit. Taken from \cite{atlas_wprime}.}
\label{fig:wprime_spectra}
\end{figure}

\section{Search for Resonances in Single Lepton Channels}

The ATLAS search for resonances in single lepton channels~\cite{atlas_wprime} selects events with exactly one electron or muon, and significant missing transverse energy~($E_{\mathrm{T}}^{\mathrm{miss}}$). The thresholds on the charged lepton transverse momentum ($p_{\mathrm{T}}$) and on $E_{\mathrm{T}}^{\mathrm{miss}}$ are of 65 and 55~GeV, respectively in the electron and muon channels. Events with more than one reconstructed charged lepton are vetoed. The lepton and the missing energy are then combined in the definition of the discovery variable, the transverse mass $m_{\mathrm{T}} = \sqrt{ 2p_{\mathrm{T}}E_{\mathrm{T}}^{\mathrm{miss}}(1-\cos\Delta\alpha)}$, where $\Delta\alpha$ is the angle, in the transverse plane, between the direction of the lepton momentum and the missing energy.

The main background to this search comes from the irreducible charged Drell-Yan production of off-shell, high-mass W bosons ($\mathrm{W^*}\rightarrow\ell\nu$, where $\ell=\mathrm{e},\mu$). This background is estimated from NLO simulation, to which is applied a mass-dependent scaling to take into account NLO-to-NNLO QCD corrections, as well as NLO electroweak corrections. 

A sub-dominant contribution to backgrounds comes from events in which the results of colored haronization emulate a reconstructed lepton (`fake' background). These events affect mainly the electron channel and are measured directly in data, by defining a control region with relaxed electron identification criteria. 


The obtained $m_{\mathrm{T}}$ spectra are shown in Figure~\ref{fig:wprime_spectra}, in the electron~(left) and muon~(right) channels. The markers represent the observed spectra on the full 36.1~fb$^{-1}$ dataset, and they are compared to the background estimates which are stacked into a histogram. Overlaid are shown also the expected shapes for benchmark Sequential Standard Model (SSM) W$^{\prime}$ signals of 3, 4, and 5~TeV mass. As can be seen from the bottom pad, where the ratio of the data to the total background estimates before~(upper pad) and especially after~(lower pad) the background fit, no significant excess is observed.

\begin{figure}[t]
\centering
\includegraphics[width=0.45\textwidth]{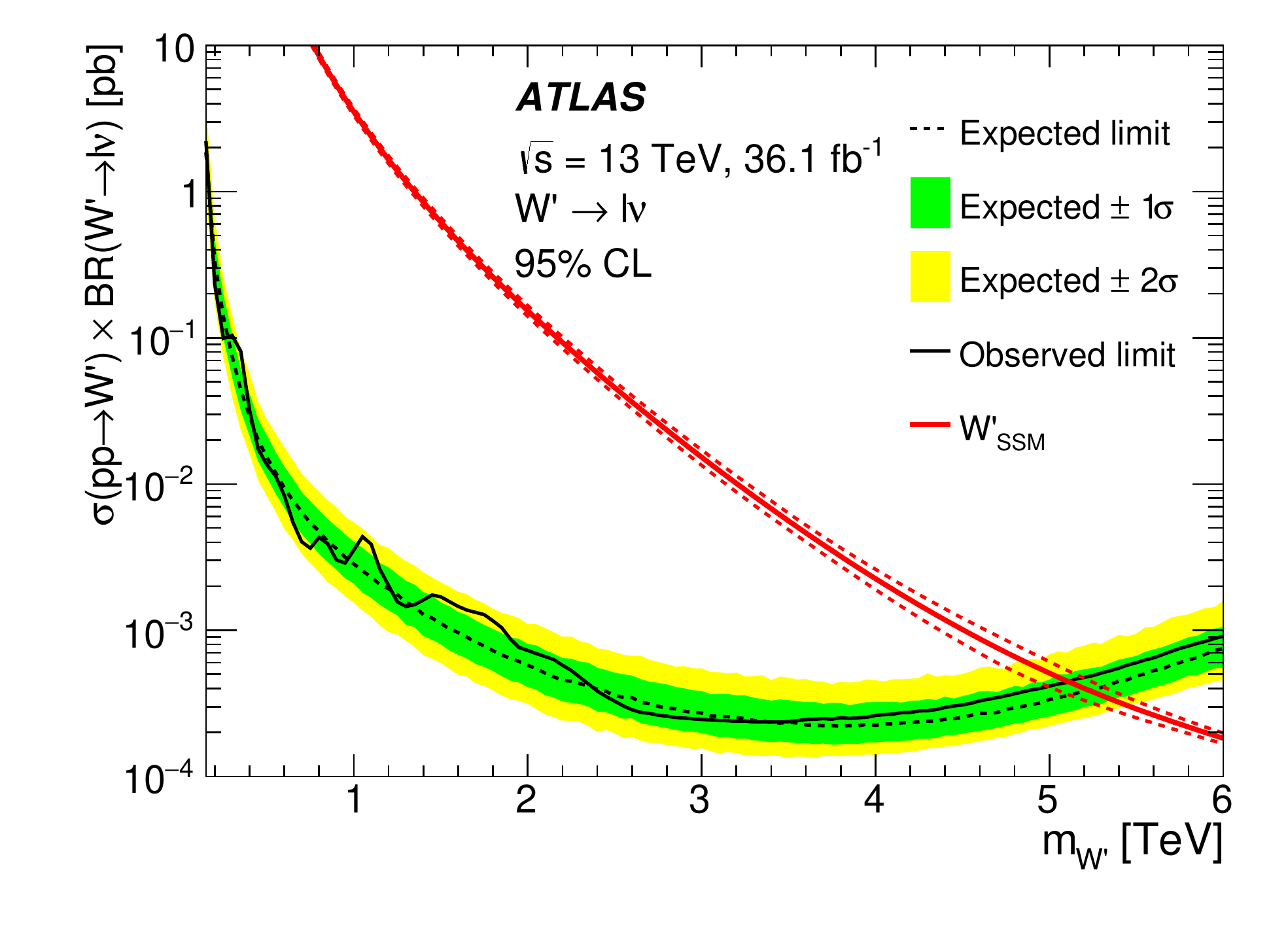}
\includegraphics[width=0.45\textwidth]{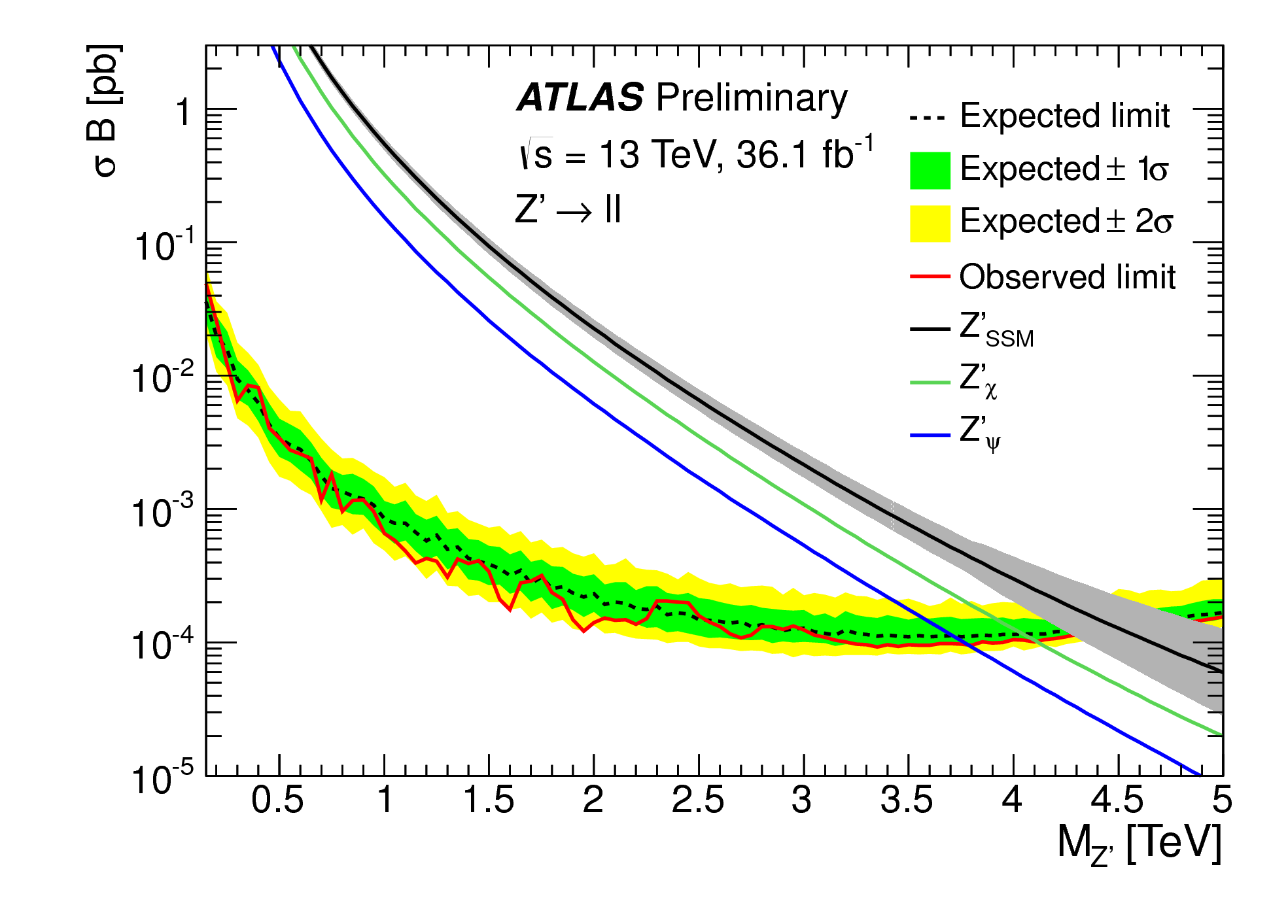}
\caption{Upper limit, at 95\% confidence level, on the production cross section of W$^{\prime}$~(left, from~\cite{atlas_wprime}) and Z$^{\prime}$~(right, from~\cite{atlas_zprime}), times the branching ratio to final states with electrons or muons, as a function of the signal particle mass. The dotted black line represents the expected upper limit, and the green and yellow bands its 68\% and 95\% uncertainty intervals, respectively. The observed limit is shown as a solid line. The expected cross section for selected W$^{\prime}$ and Z$^{\prime}$ models are overlaid. }
\label{fig:wprime_limit}
\end{figure}

Upper limits are set on the production cross section of W$^{\prime}$ particles, which can be seen in Figure~\ref{fig:wprime_limit}~(left): the dotted black line represents the expected upper limit at 95\% confidence level, and the green and yellow bands its 68\% and 95\% uncertainty intervals, respectively. The observed limit is shown as a solid black line. As a comparison, the cross section for SSM W$^{\prime}$ is shown in red: as can be seen this search excludes this model up to a mass of 5.1~TeV. Similar results were published by CMS on a dataset of 2.3~fb$^{-1}$, with which masses up to 4.1~TeV were excluded~\cite{cms_wprime}.

\section{Search for Resonances in Dilepton Channels}

The ATLAS search for resonances in dilepton channels~\cite{atlas_zprime} selects events with two same-flavor charged leptons, either electrons or muons, with $p_{\mathrm{T}}>30$~GeV. The invariant mass of the dilepton system is the discovery variable, and it is scanned in the high-mass region to search for resonances.

The main background to this search is Drell-Yan production of off-shell high-mass Z bosons, and it is estimated through NLO simulation. Similarly to the single lepton search, mass-dependent scale factor are applied, to take into account higher-order QCD and electroweak corrections. Small contributions from other processes which also have two genuine leptons in the final state (such as events with top quark or heavy boson pairs) are also taken from the simulation. A small contribution to the background, which mainly affects the electron channel, comes from events with jets faking leptons. This is estimated in data with a method similar to the one used in the single lepton channel, adapted to the case of two leptons in the final state. 

\begin{figure}[htb]
\centering
\includegraphics[width=0.42\textwidth]{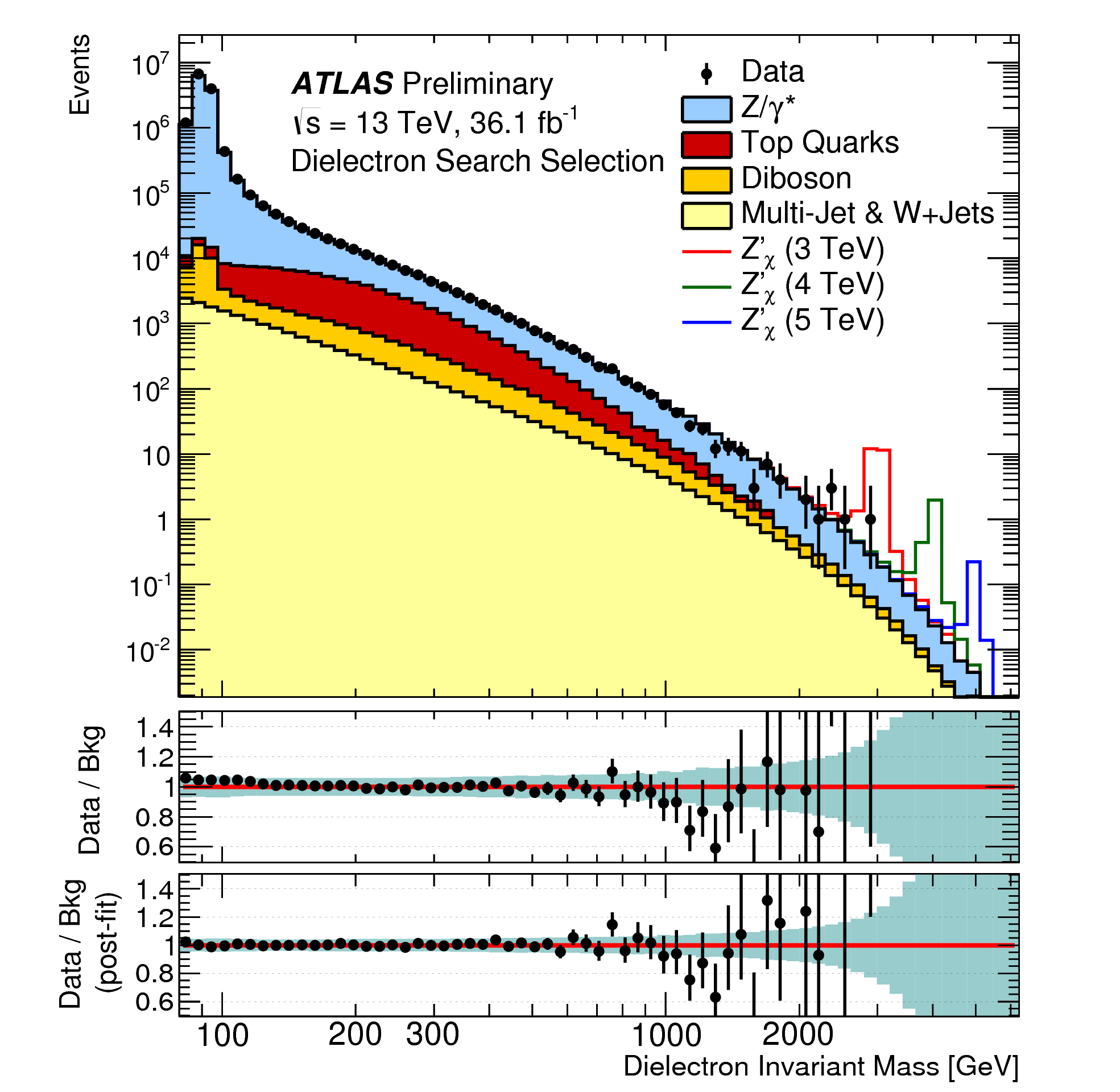}
\includegraphics[width=0.42\textwidth]{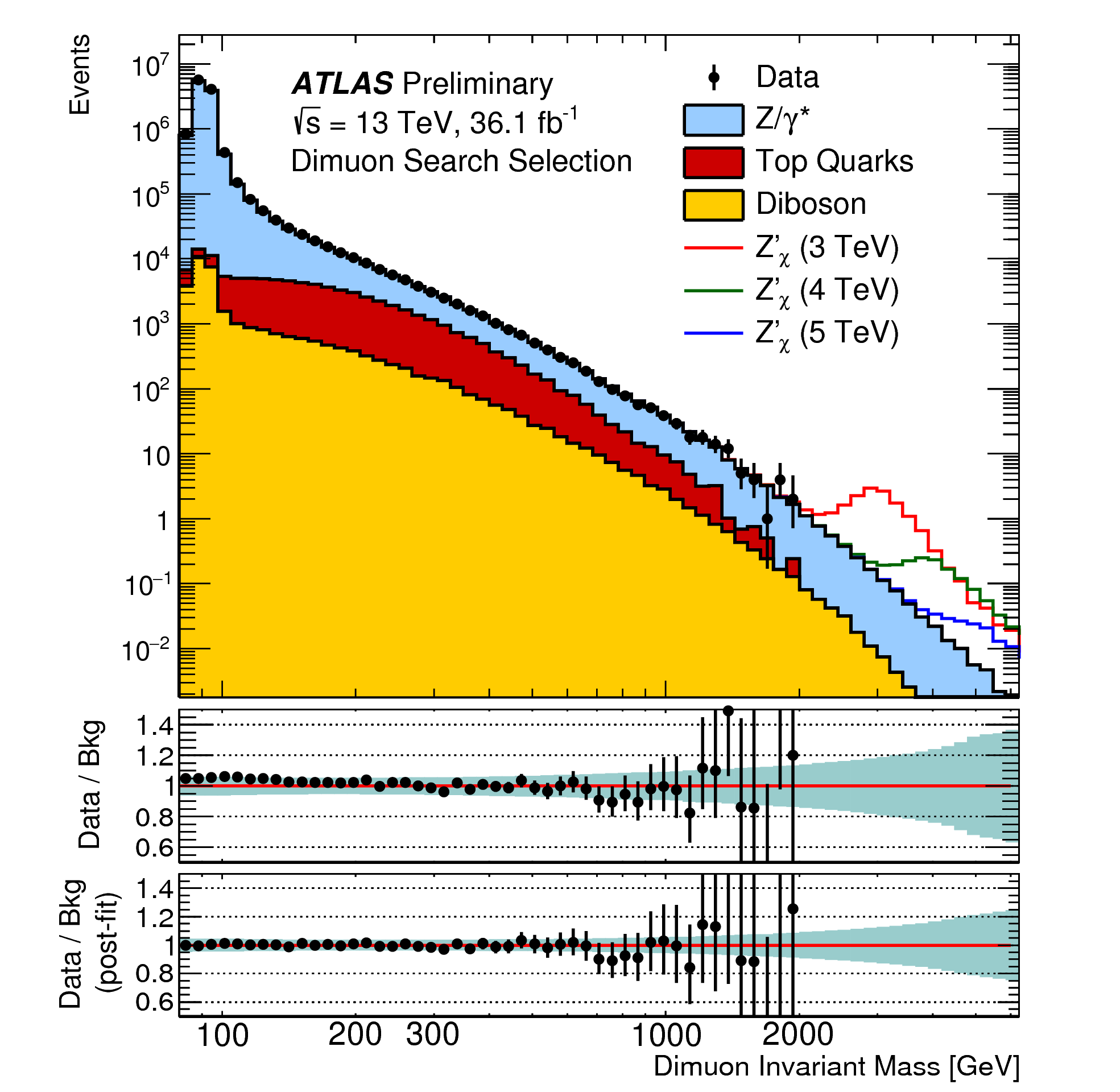}
\caption{Transverse mass spectra of the ATLAS dilepton searches in the electron~(left) and muon~(right) channels. The markers represent the observed spectra on the full 36.1~fb$^{-1}$ dataset, while the stacked histograms represent the background estimates. Overlaid are shown also the expected shapes for benchmark Z$^{\prime}$ signals of 3, 4, and 5~TeV mass. The bottom pads show the ratio of the data to the sum of all backgrounds before (upper pad) and after~(lower pad) the background fit. Taken from \cite{atlas_zprime}.}
\label{fig:zprime_spectra}
\end{figure}


The invariant mass spectra of this search are found in Figure~\ref{fig:zprime_spectra}, for the electron~(left) and muon~(right) channels. The data, corresponding to 36.1~fb$^{-1}$, is shown with black markers, and is compared to the stacked background estimates. As can be seen from the bottom pads, no significant excess is observed. The corresponding 95\% confidence level upper limits on the production of Z$^{\prime}$ signals are shown in Figure~\ref{fig:wprime_limit}~(right). The cross sections for three benchmark Z$^{\prime}$ models (SSM, Z$^{\prime}_{\chi}$, Z$^{\prime}_{\psi}$) are also shown. As can be seen this analysis excludes Z$^{\prime}_{\mathrm{SSM}}$ below 4.5~TeV, and Z$^{\prime}_{\psi}$ below 3.8~TeV. A similar analysis conducted by CMS~\cite{cms_zprime} on 13~fb$^{-1}$ of data excludes those same models for masses below 4.0 and 3.5~TeV, respectively.

\section{Search for Resonances in Diphoton Channels}

The CMS search for diphoton resonances~\cite{cms_gg} selects events with two photons with $p_{\mathrm{T}} > 75$~GeV, at least one of which with pseudorapidity $|\eta|<1.4$. The events are then categorized into the EBEB and EBEE classes, respectively if both or only one of the photons have $|\eta|<1.4$. The continuous background in this search is estimated directly in the data, through a fit of the diphoton mass spectrum. The background shape is parametrized with the function $f(x) = x^{a+b\log x}$, and the uncertainty related to the choice of this particular functional form was assessed through bias studies.

\begin{figure}[htb]
\centering
\includegraphics[width=0.4\textwidth]{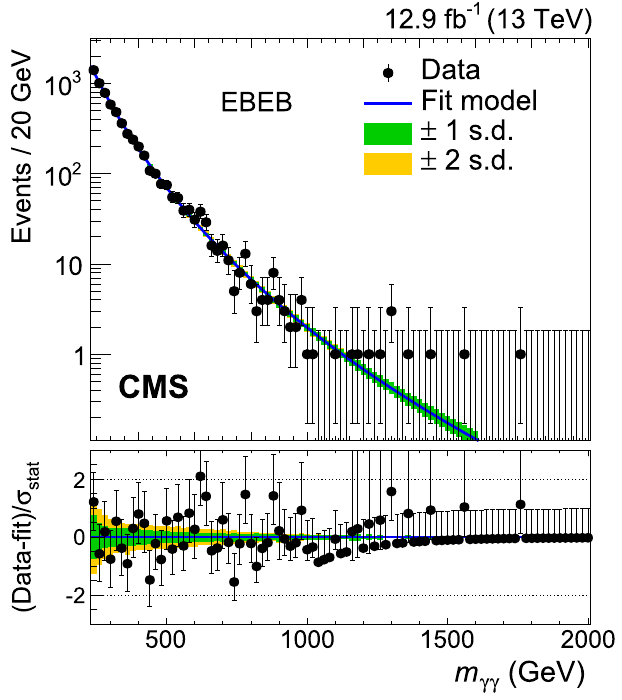}
\includegraphics[width=0.4\textwidth]{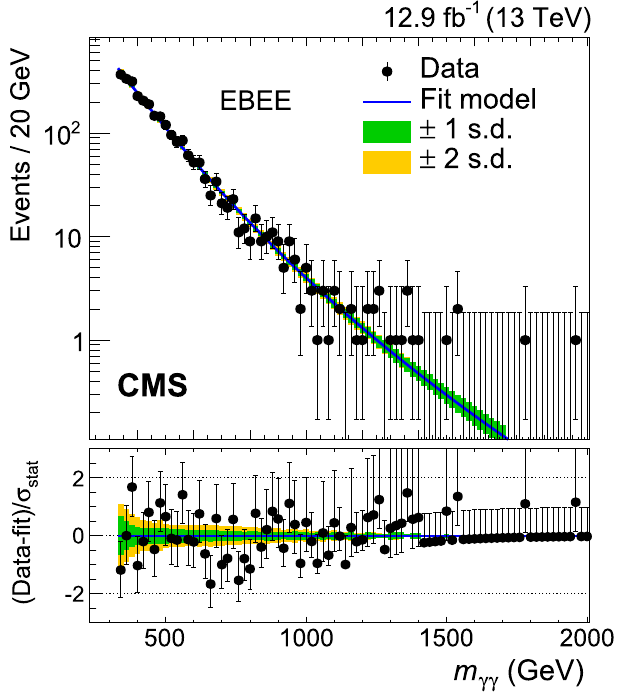}
\caption{Invariant mass spectra of the CMS diphoton search, for the EBEB~(left) and EBEE~(right) categories. The background fit is shown with a blue line, together with its 68\%~(green) and 95\%~(yellow) uncertainty bands. The bottom pad shows the difference between the data and the fit, divided by the statistical uncertainty. Taken from \cite{cms_gg}.}
\label{fig:gg_mass}
\end{figure}

Figure~\ref{fig:gg_mass} shows the invariant mass spectra obtained in 12.9~fb$^{-1}$ of data, in the EBEB~(left) and EBEE~(right) channels. The fitted background shapes are shown as blue lines, together with their 68\%~(green) and 95\%~(yellow) uncertainty bands. The ratio between the fit and the data, divided by the statistical uncertainty, is shown in the bottom pad. 

\begin{figure}[htb]
\centering
\includegraphics[width=0.45\textwidth]{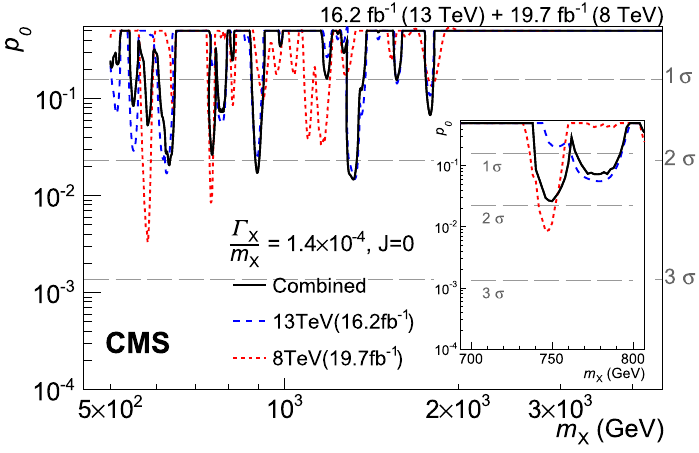}
\includegraphics[width=0.45\textwidth]{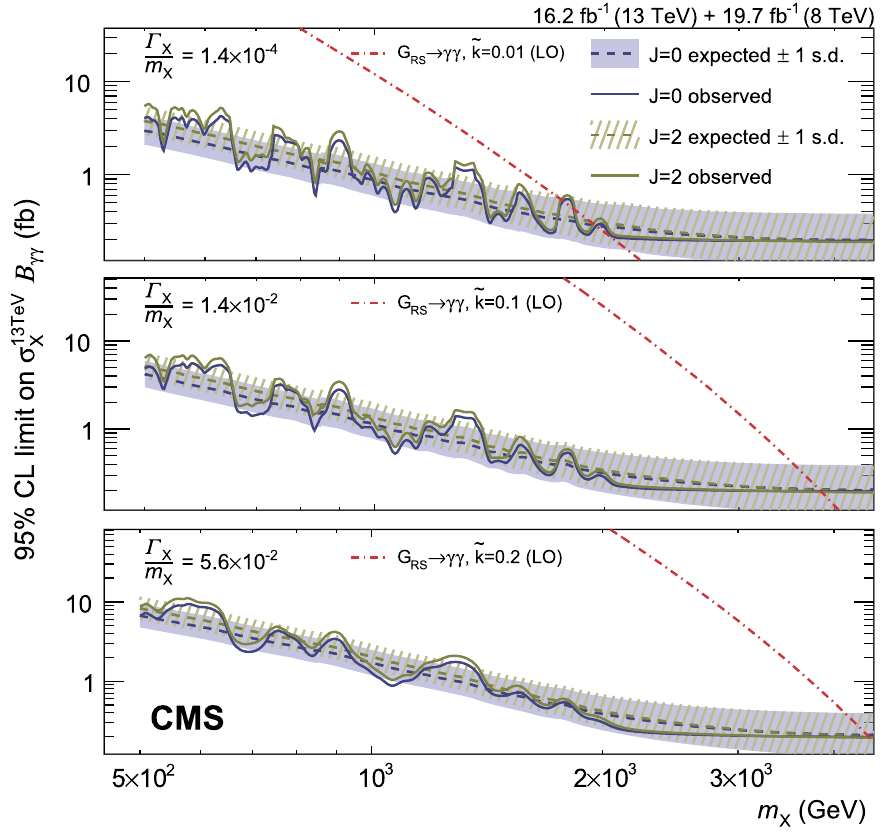}
\caption{Left: observed p-values of the background-only hypothesis, as a function of mass, for the 8~TeV dataset~(red dotted), 13~TeV dataset~(blue dashed) and their combination~(black, solid). Right: 95\% upper limits on the production cross section of spin-0 and spin-2 resonances, for three differents widths: 0.014\%~(upper), 1.4\%~(middle), 5.6\%~(lower). The expected cross section for Randall-Sundrum gravitons is shown with a red dashed line. Taken from \cite{cms_gg}.}
\label{fig:gg_limit}
\end{figure}

The results on this dataset are combined with previously-collected 13 TeV data, and with the full 8~TeV dataset, to produce the results shown in Figure~\ref{fig:gg_limit}. The left plot shows the observed $p$-values of the background-only hypothesis: the red line corresponds to the 8~TeV dataset, the blue one to 13~TeV, and the black one shows their combination. As can be seen, no significant excess is observed. The right plot in Figure~\ref{fig:gg_limit} shows 95\% confidence level upper limits on the production of spin-0 or spin-2 resonances, as a function of their mass, for three possible resonance widths. The expected cross section for a Randall-Sundrum graviton is overlaid with a red line, and as can be seen this search excludes the presence of such particles between 1.95 and 4.45~TeV, depending on the width. Similar results have been produced by the corresponding ATLAS diphoton resonance search~\cite{atlas_gg}.

\section{Search for $Z\gamma$ Resonances}

The CMS search for $Z\gamma$ resonances~\cite{cms_zg} is conceptually similar to the diphoton search, but adds sensitivity to spin-1 particles. The search is split into two broad categories, depending on the decay of the Z boson: the leptonic channel, which is characterized by high resolution and low backgrounds, and is further subdivided into the ee$\gamma$ and $\mu\mu\gamma$ categories; and the hadronic channel, which has lower resolution and higher backgrounds, but also higher signal branching ratio, which pays off in the high-mass background-free region. Because this is a high-mass resonance search, the Z boson produced in signal decays will have a strong Lorentz boost: for this reason the hadronic channel uses large cones for jet reconstruction, and further categorized events based on jet substructure variables such as the `subjettiness' variable $\tau_{21}$, and a novel b-tagging algorithm based on the jet {\em subjets}.

\begin{figure}[htb]
\centering
\includegraphics[width=0.32\textwidth]{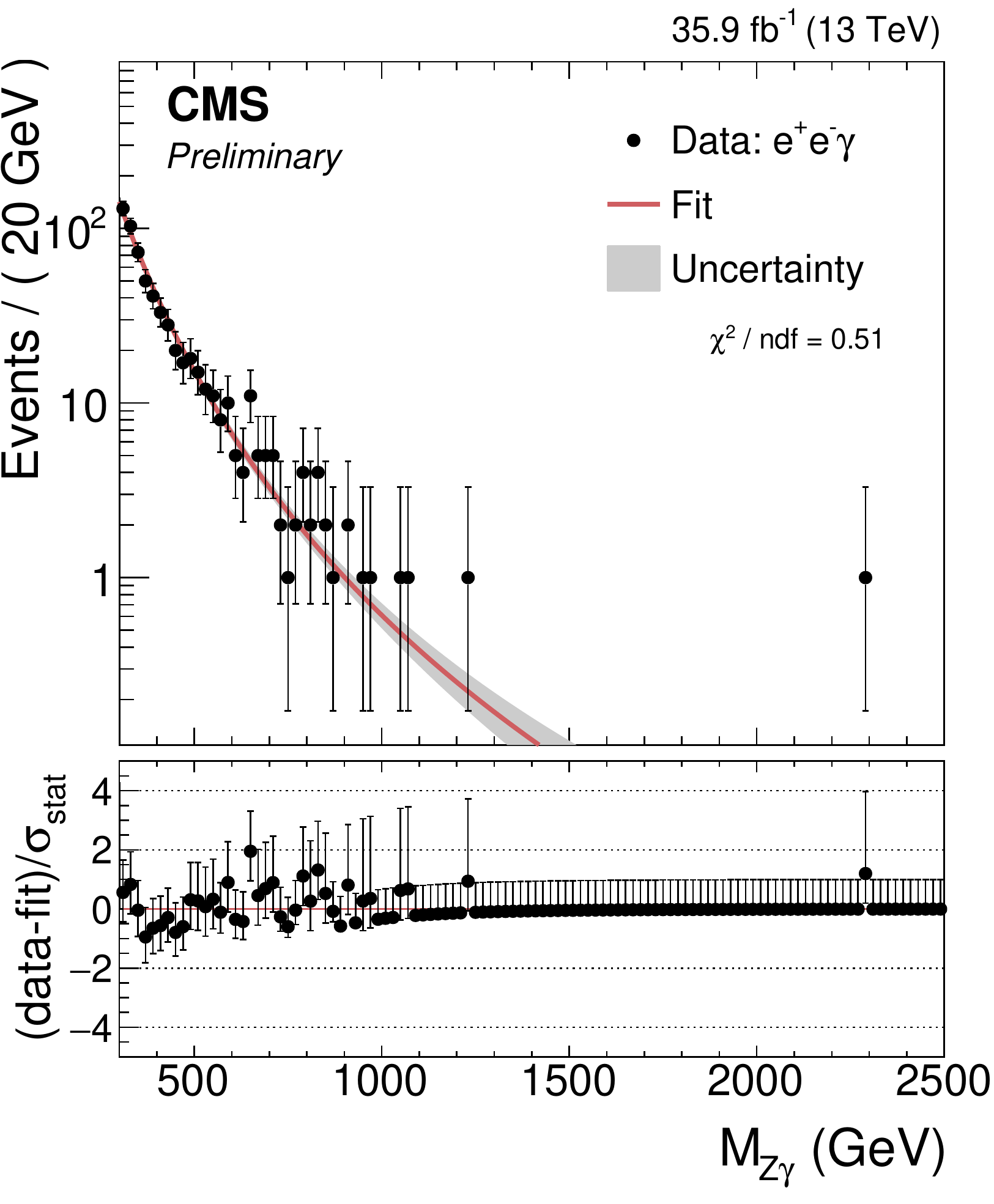}
\includegraphics[width=0.32\textwidth]{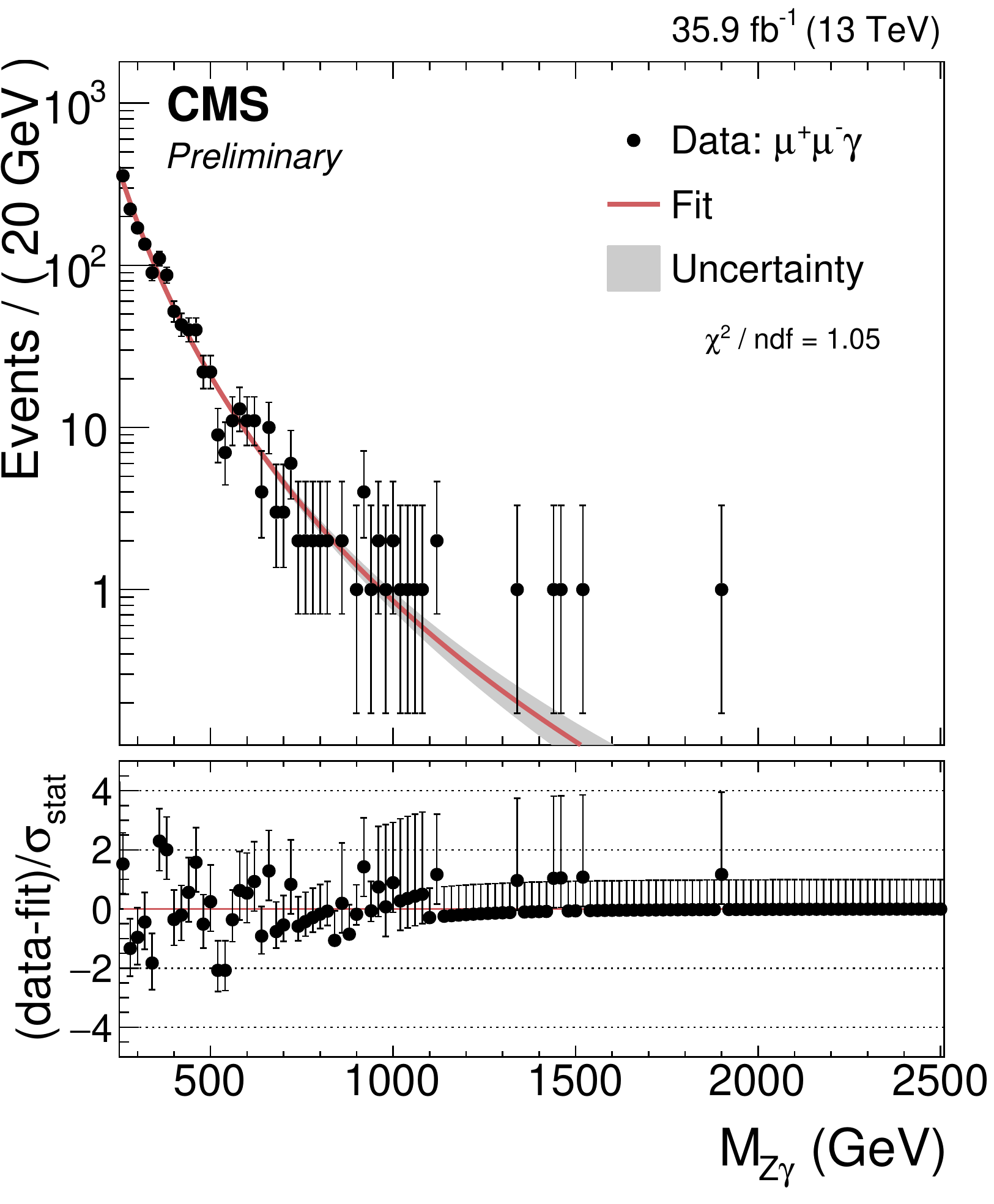}
\includegraphics[width=0.32\textwidth]{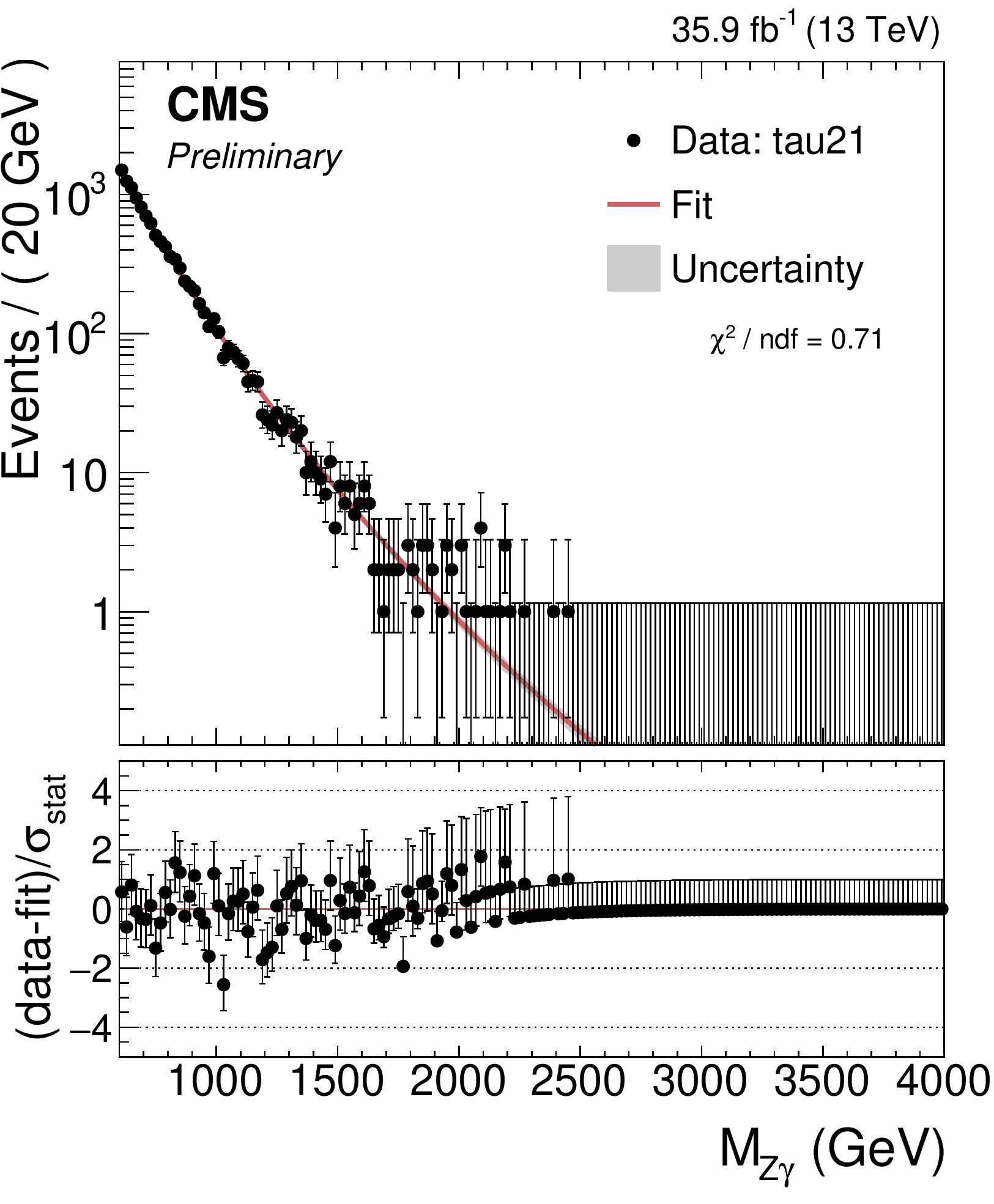}
\caption{Invariant mass spectra of the Z$\gamma$ analysis, in the ee$\gamma$~(left), $\mu\mu\gamma$~(center) and a selected hadronic category~(right). The results of the background fits are shown with a solid line, together with its 68\%~(green) and 95\%~(yellow) uncertainty bands. The bottom pad shows the difference between the data and the fit, divided by the statistical uncertainty. Taken from \cite{cms_zg}.}
\label{fig:zg_mass}
\end{figure}

The invariant mass spectra obtained on 35.9~fb$^{-1}$ of data are shown in Figure~\ref{fig:zg_mass}: the left plot is the ee$\gamma$ channel, the middle one is $\mu\mu\gamma$ and the right plot is a selected category of the hadronic analysis. In all categories the background is parametrized with the same functional form used in the diphoton analysis described above, and the result of the fits are shown with a red line and a grey uncertainty band. The bottom pads show the ratios between the data and the fit, divided by the statistical uncertainty.

\begin{figure}[htb]
\centering
\includegraphics[width=0.32\textwidth]{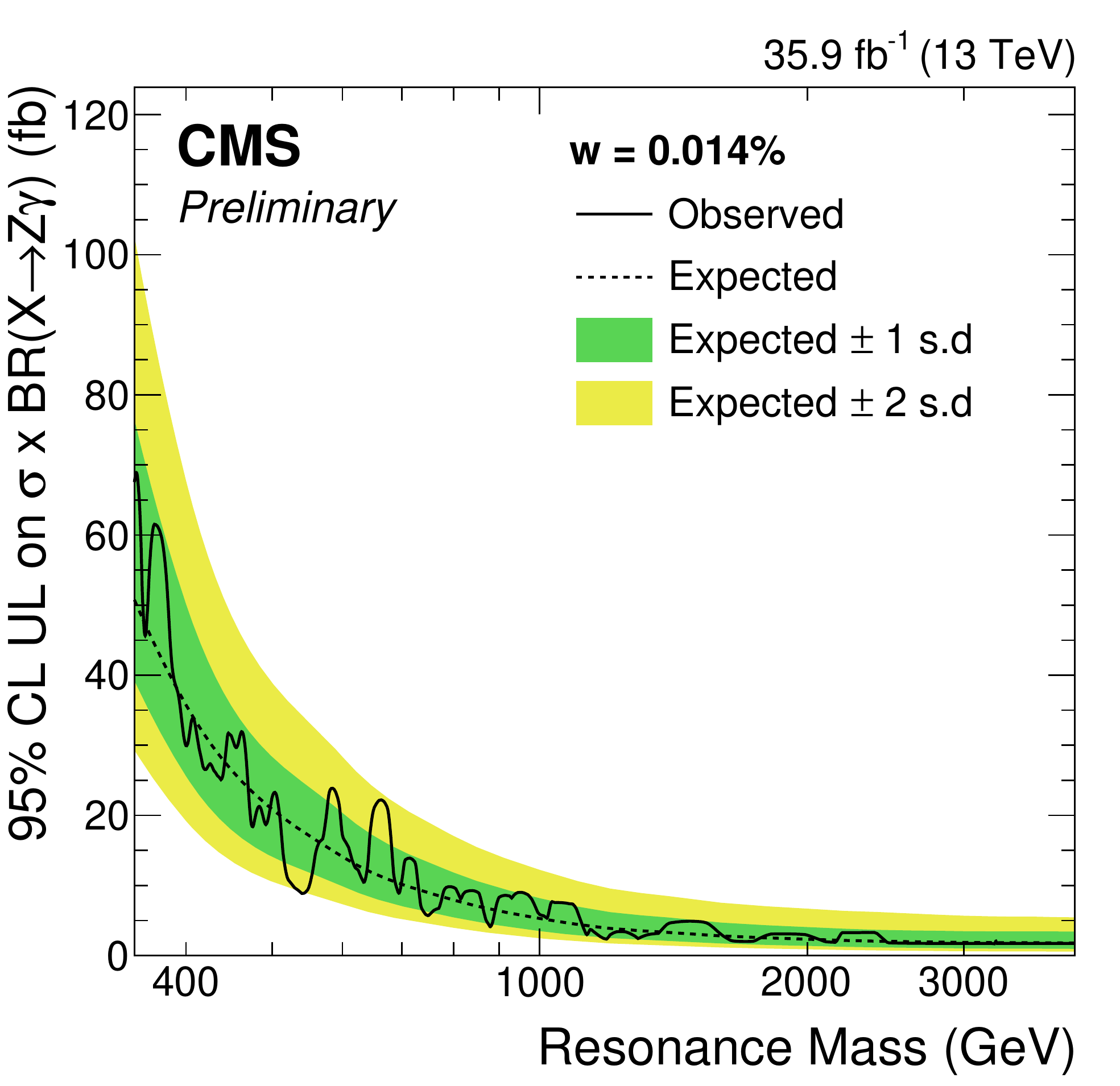}
\includegraphics[width=0.32\textwidth]{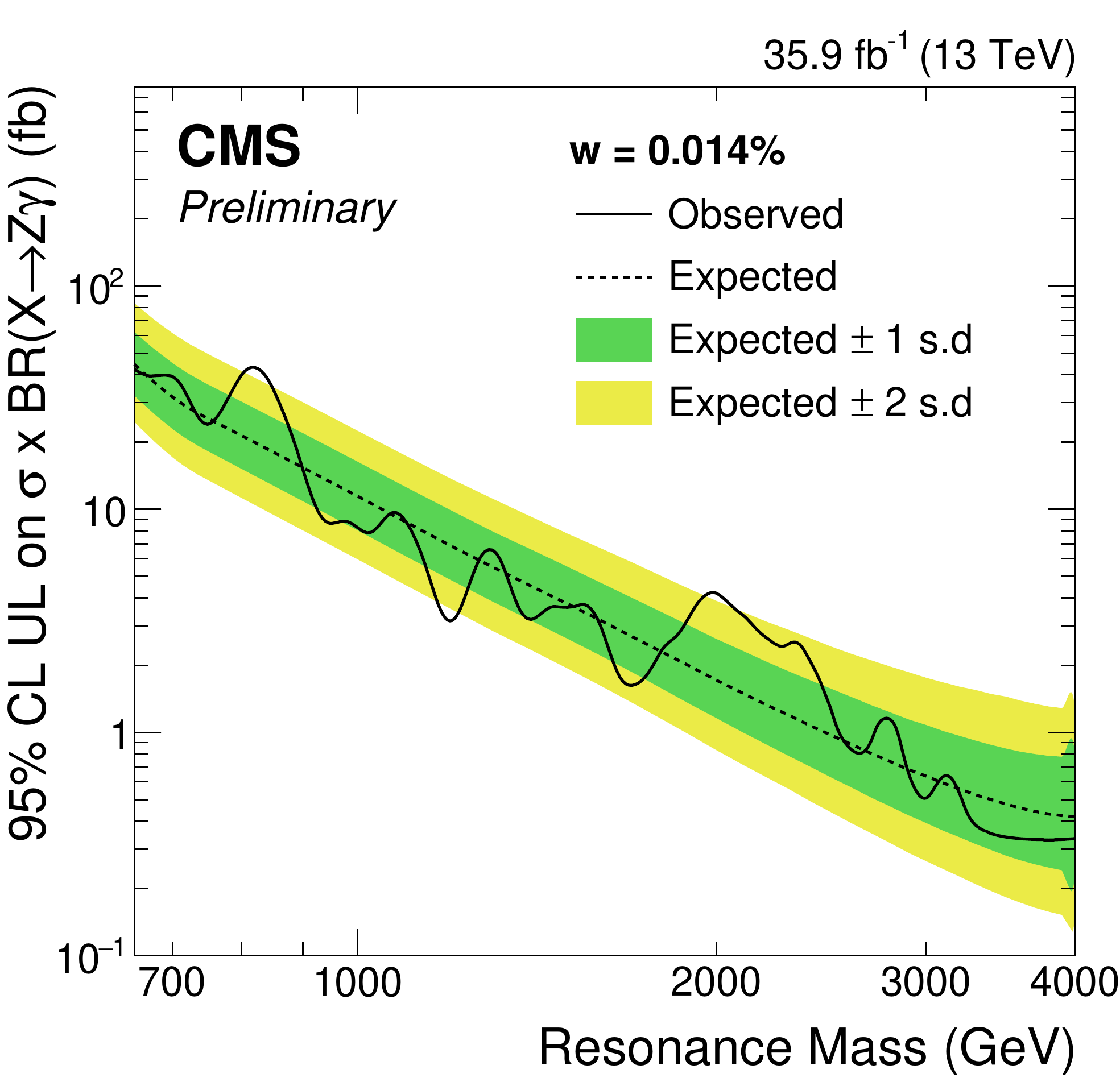}
\includegraphics[width=0.32\textwidth]{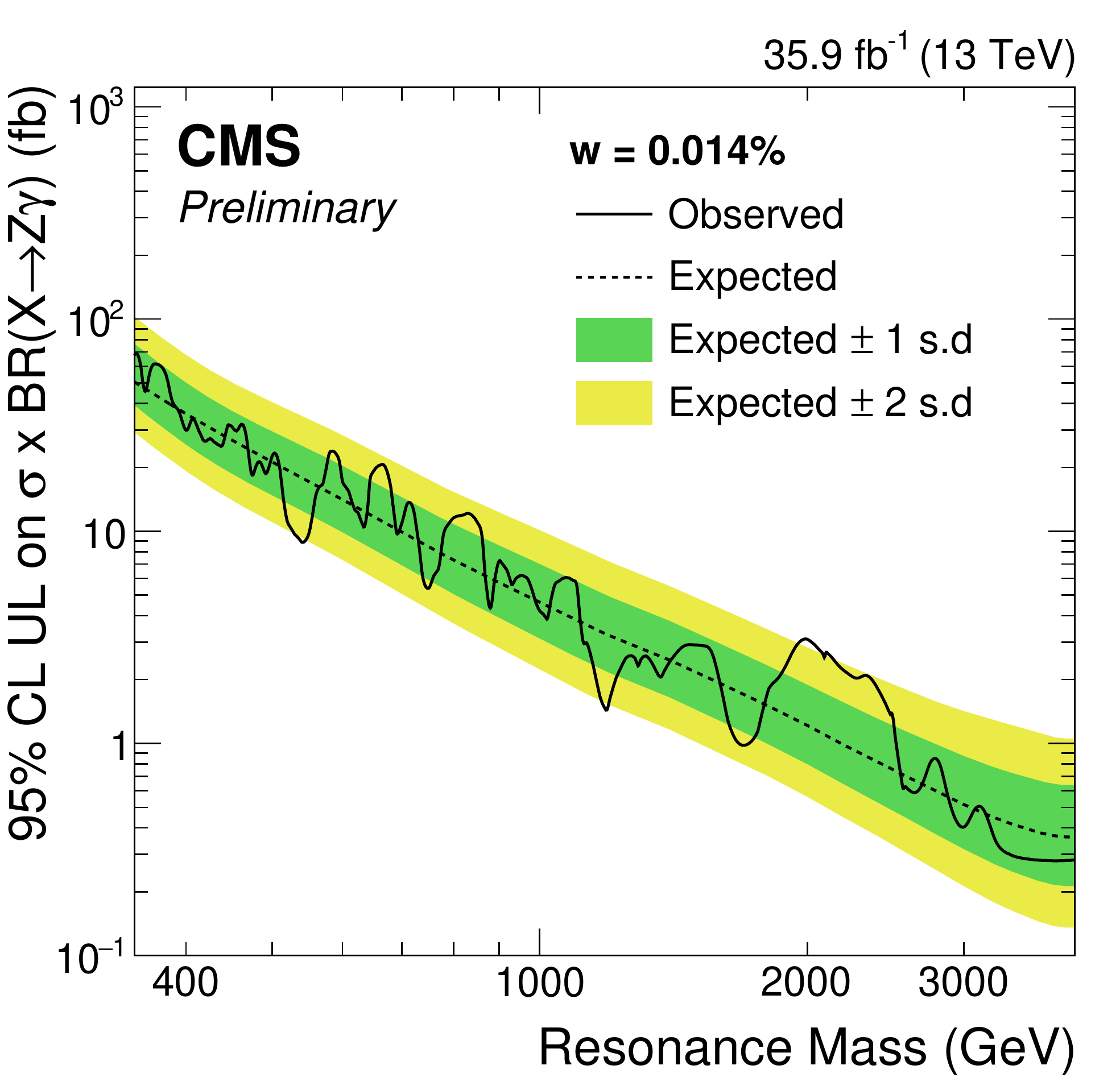}
\caption{Upper limit, at 95\% confidence level, on the production cross section of Z$\gamma$ resonances, as a function of the resonance mass, for the leptonic channel~(left), the hadronic channel~(center) and their combination~(right). The dotted black line represents the expected upper limit, and the green and yellow bands its 68\% and 95\% uncertainty intervals, respectively. The observed limit is shown as a solid black line. Taken from \cite{cms_zg}. }
\label{fig:zg_limit}
\end{figure}

Figure~\ref{fig:zg_limit} shows the 95\% confidence level upper limits on the production cross section of narrow resonances, as a function of their mass, in the leptonic channel~(left), in the hadronic channel~(middle) and in their combination~(right). The observed limit~(solid line) is compared to the expected one~(dashed line) and its uncertainty bands. As can be seen no significant deviation is observed from the background-only hypothesis. Similar results have been produced by the ATLAS Z$\gamma$ search~\cite{atlas_zg}.

\section{Conclusions}

Searches for lepton and photon resonance have always been a powerful tool for discovery in high energy physics. We have presented the latest results on such searches performed on up to 36.1~fb$^{-1}$ of 13~TeV proton-proton collisioned produced in the LHC and analyzed by the ATLAS and CMS experiments. No significant excess have been observed in the single lepton, dilepton, diphoton and Z$\gamma$ searches, but the reach of these analyses has been significantly extended with respect to previously published results.


\end{document}

%% file: econfmacros.tex
\def\beq{\begin{equation}}
\def\eeq#1{\label{#1}\end{equation}}
\def\eeqn{\end{equation}}

\def\beqa{\begin{eqnarray}}
\def\eeqa#1{\label{#1}\end{eqnarray}}
\def\eeqan{\end{eqnarray}}

\let\bar=\overbar

\def\Dslash{\not{\hbox{\kern-4pt $D$}}}
\def\dslash{\not{\hbox{\kern-2pt $\del$}}}

\def\msb{{\bar{\ssstyle M \kern -1pt S}}}

